# Singling out people without knowing their names - Behavioural targeting, pseudonymous data, and the New Data Protection Regulation

Frederik J. Zuiderveen Borgesius[1]

*Pre-print. Published version:*

## Table of Contents

[1] Dr. Frederik Zuiderveen Borgesius, researcher at Institute for Information Law (IViR), University of Amsterdam, The Netherlands. Email address: F.J.ZuiderveenBorgesius@uva.nl.
This paper is based on, and includes sentences from, the author's PhD thesis *Improving Privacy Protection in the area of Behavioural Targeting* (Kluwer Law International 2015). The author wishes to thank Solon Barocas, Ian Brown, David Erdos, Nico van Eijk, Dirk Henderickx, Joris van Hoboken, Matthijs Koot, Manon Oostveen, Corrette Ploem, João Quintais, Rachel Wouda, and the anonymous reviewers for their helpful comments. Any errors are the author's own.

**Abstract**

Information about millions of people is collected for behavioural targeting, a type of marketing that involves tracking people's online behaviour for targeted advertising. It is hotly debated whether data protection law applies to behavioural targeting. Many behavioural targeting companies say that, as long as they do not tie names to data they hold about individuals, they do not process any personal data, and that, therefore, data protection law does not apply to them. European Data Protection Authorities, however, take the view that a company processes personal data if it uses data to single out a person, even if it cannot tie a name to these data. This paper argues that data protection law should indeed apply to behavioural targeting. Companies can often tie a name to nameless data about individuals. Furthermore, behavioural targeting relies on collecting information about individuals, singling out individuals, and targeting ads to individuals. Many privacy risks remain, regardless of whether companies tie a name to the information they hold about a person. A name is merely one of the identifiers that can be tied to data about a person, and it is not even the most practical identifier for behavioural targeting. Seeing data used to single out a person as personal data fits the rationale for data protection law: protecting fairness and privacy.

**Keywords**

## 1    Introduction

It is hotly debated whether data protection law applies to behavioural targeting. Behavioural targeting, or online profiling, is a type of personalised communication that involves monitoring people's online behaviour and using the collected information to show people individually targeted advertisements. Many behavioural targeting companies say that, as long as they do not tie names to data they hold about individuals, they do not process any personal data, and that, therefore, data protection law does not apply to them. This paper examines whether data protection applies to behavioural targeting, and whether, from a fundamental rights perspective, it should apply.

Behavioural targeting and data protection law are introduced in Section 2 and 3. In Section 4, it is shown that, from a doctrinal perspective, nameless data can be viewed as personal data when a company uses these data to *single out* a person, a view taken by European Data Protection Authorities. Apart from that, section 5 explains that it is often fairly easy to tie a name to behavioural targeting data. The new Data Protection Regulation and its definitions of 'personal data' and 'pseudonymous data' are discussed in Section 6. Section 7 argues that data protection law should apply to behavioural targeting. Counter-arguments are considered in Section 8. The conclusion is provided in Section 9: data protection law generally applies – and should apply – to behavioural targeting.

## 2    Targeted online marketing

Information about millions of people is collected for behavioural targeting. For instance, Facebook collects information about at least 1.5 billion people.[2] Google says it 'reaches 90% of Internet users worldwide.'[3] Some lesser-known companies also process information about many people, such as the Rubicon Project ('600 million'),[4] and AddThis ('1.9 billion').[5]

---

[2] Facebook says it had '1.55 billion monthly active users as of September 30, 2015' <http://newsroom.fb.com/company-info/> accessed 24 January 2016.
[3] Google Adwords, 'About the Google Display Network' (publication date unknown) <https://adwords.google.com/support/aw/bin/answer.py?hl=en&answer=57174> accessed 24 January 2016.

Many types of companies are involved with behavioural targeting, and the resulting data flows are complicated. In a simplified example of behavioural targeting, an advertising network follows an internet user's behaviour, so it can display individually targeted ads to this user. Ad networks are companies that serve advertisements on thousands of websites. An ad network can track a person's visits to all websites on which it serves ads.

Ad networks often use cookies. These are small text files that website publishers can store on an Internet user's computer. If the cookie contains a unique identifier, website publishers can recognise the visitor's computer. Recognising a visitor's computer is useful, for instance if somebody has included items in a virtual shopping cart. Another example is that of language selected on a website, after which the website publisher can store a cookie on the visitor's computer to ensure that the website will be displayed in the selected language at every subsequent visit by the same individual.

There are several types of cookies. Session cookies are deleted when the user closes his or her browser. Persistent cookies are retained when the user closes the browser or turns off the computer. First party cookies are placed by website publishers. Third party cookies are placed through a website by other parties than the website publisher. Tracking cookies that are used to recognise people contain unique codes, such as 22be6e056ca010062||t=1392841778|cs=002213fd48e6bd6f7bf8d99065.[6] If a website publisher uses a cookie to remember a visitor's language settings, the publisher can use a cookie without a unique identifier, for instance FR for French, or EN for English.

When visiting a website, say Newspaper.com, it seems like all parts of the website are presented by one publisher. In reality, various elements on a website are often presented by different companies. The widget of www.newspaper.com showing the weather report might be served from www.weather.com. If ads are displayed on www.newspaper.com, these might be served from www.adnetwork.com, and a Facebook 'Like' button on a website is served by Facebook.[7] All these third parties can store and read their own cookies. During a single website visit, the visitor may receive dozens of third-party tracking cookies.[8]

If www.newspaper.com stores a cookie on a computer, in principle other websites, such as www.gossip.com, cannot read that cookie. Hence, in principle websites can only read their own cookies. However, ad networks have found a way around this system. Third parties such as Weather.com and Adnetwork.com can set and read their own cookies. Hence, Adnetwork.com can set and read its cookies through www.newspaper.com and www.gossip.com, if it serves ads on both websites. In this way, Adnetwork.com can recognise visitors on any website on which it serves advertising. Third-party cookies that are used to follow people around the web are referred to as tracking cookies. The Interactive Advertising Bureau, a trade association for online and mobile advertising, explains that 'cookies are used in behavioural advertising to identify users who share a particular interest so that they can be served more relevant adverts.'[9]

Apart from cookies, behavioural targeting companies use many other tracking technologies. Some technologies, such as flash cookies and other super cookies, are comparable to conventional cookies and involve storing a unique identifier on devices. While people can delete conventional cookies from their computers, super cookies are usually harder to delete. Some companies have used flash cookies to reinstall, or re-spawn, cookies that people deleted: 'zombie cookies'.[10]

[7] Güneş Acar, Brendan Van Alsenoy, Frank Piessens, Claudia Diaz and Bart Preneel, 'Facebook Tracking Through Social Plug-ins' (Technical report prepared for the Belgian Privacy Commission) (V. 1.1, 24 June 2015) <https://securehomes.esat.kuleuven.be/~gacar/fb_tracking/fb_plugins.pdf> accessed 24 January 2016.
[8] Chris Jay Hoofnagle and Nathan Good, Web Privacy Census (1 June 2012) <http://ssrn.com/abstract=2460547> accessed 24 January 2016.
[9] Interactive Advertising Bureau, 'A Guide to online behavioural advertising' (Internet marketing handbook series) (2009) <www.iabuk.net/sites/default/files/publication-download/OnlineBehaviouralAdvertisingHandbook_5455.pdf> accessed 24 January 2016, p. 4.

Other tracking methods do not rely on storing an identifier on a device. For example, passive device fingerprinting involves recognising a device by analysing the information it transmits. A computer's browser can be recognised by looking at characteristics such as the browser type (e.g. Mozilla Firefox version 38.0.5), its settings, and installed fonts. A device fingerprint is 'a set of system attributes that, for each device, take a combination of values that is, with high likelihood, unique, and can thus function as a device identifier.'[11] Some companies use device fingerprinting for behavioural targeting. One company claims to have fingerprinted 3 billion devices.[12] While some savvy users may know how to delete flash cookies and other identifiers, it is very difficult to prevent a device from being fingerprinted. Although there are many tracking technologies, for readability reasons, this paper mostly refers to cookies.

Information captured for behavioural targeting can concern many online activities: what people read, which videos they watch, what they search for, etc. Companies can enrich individual profiles with data gathered online and offline. Some companies can collect up-to-date location data of users' mobile devices, showing where the user is in almost real-time.[13] Various parties have access to such location data. Smart phone apps sometimes send location data to the app provider, or to ad networks.[14] Companies can track people's movements in shops by analysing signals emitted by people's phones, such as Bluetooth and Wi-Fi signals.[15] Some marketers have high hopes for advertising on mobile devices.[16] If somebody's profile suggests that she likes sushi, a Japanese restaurant might advertise a deal when she is in the area around lunchtime.

---

[10] Christian Olsen, 'Supercookies: What You Need to Know About the Web's Latest Tracking Devic' (Mashable) (2 September 2011) <http://mashable.com/2011/09/02/supercookies-internet-privacy/> accessed 24 January 2016. See also Mika D Ayenson, Nathaniel Good, Chris Jay Hoofnagle, Ashkan Soltani, Dietrich J. Wambach, 'Behavioral advertising: The offer you cannot refuse' (2012) *Harvard Law & Policy Review, 6*(2), 273-296.

[11] Günes Acar, Marc Juarez, Nick Nikiforakis, Claudia Diaz, Seda Gürses, Frank Piessens and Bart Preneel, 'FPDetective: Dusting the web for fingerprinters' (2013) CCS '13 Proceedings of the 2013 ACM SIGSAC conference on Computer & Communications Security 1129.

[12] 'iovation has intelligence on over 3 billion global devices and that number grows every day.' <www.iovation.com> accessed 24 January 2016.

[13] Center for Democracy & Technology, 'Threshold Analysis for Online Advertising Practices' (January 2009) <www.cdt.org/privacy/20090128threshold.pdf> accessed 24 January 2016, p. 16.

One way to enrich user profiles is cookie synching, or cookie matching. The Interactive Advertising Bureau describes cookie synching as 'a method of enabling data appending by linking one company's user identifier to another company's user identifier, to create a richer user profile at the cookie level.'[17] Cookie synching happens routinely. Research has brought to light that the cookies of Google's DoubleClick ad network are synched with cookies of over 125 other companies.[18]

In computer science, nameless profiles of individuals, such as those used for behavioural targeting, are referred to as pseudonymous: 'A pseudonym is an identifier of a subject other than one of the subject's real names.'[19] Many behavioural targeting companies, however, say they only process 'anonymous' data and that therefore data protection law does not apply. For example, the Interactive Advertising Bureau says about behavioural

---

targeting: 'The information collected and used for this type of advertising is not personal, in that it does not identify you – the user – in the real world. No personal information, such as your name, address or email address, is used. Data about your browsing activity is collected and analysed anonymously.'[20] However, below we will see that European Data Protection Authorities say that a company processes personal data if it uses data to single out a person, regardless of whether a name can be tied to the data.

## 3    Data protection law

In Europe, data protection law is the main legal instrument to protect fairness and privacy when information about people is used.[21] Data protection law grants rights to people whose data are being processed (data subjects), and imposes obligations on parties that process personal data (data controllers, also referred to as companies in this paper).[22] The European Union Charter of Fundamental Rights states that personal 'data must be processed fairly for specified purposes and on the basis of the consent of the person concerned or some other legitimate basis laid down by law. Everyone has the right of access to data which has been collected concerning him or her, and the right to have it rectified.'[23] Independent Data Protection Authorities oversee compliance with the rules.[24] The Data Protection Directive lays down more detailed rules for personal data processing.

---

[20] Interactive Advertising Bureau Europe. Your Online Choices. A Guide to Online Behavioural Advertising, About <www.youronlinechoices.com/uk/about-behavioural-advertising> accessed 24 January 2016.

[21] The Data Protection Directive 95/46, O.J.1995, L281. See Art. 6(1) (fairness) and Art. 1(1) (privacy). The European Union Charter of Fundamental Rights also refers to fair processing (Art. 8(2)).

[22] Art. 2(a) and 2(h) of the Data Protection Directive. The directive distinguishes data controllers from data processors. That distinction falls outside the scope of this paper. See on the distinction: Brendan Van Alsenoy, 'Allocating responsibility among controllers, processors, and 'everything in between': the definition of actors and roles in Directive 95/46/EC' [2012] 28 Computer Law & Security Review 25.

[23] Art. 8(2) of the European Union Charter of Fundamental Rights.

[24] Art. 8(3) of the European Union Charter of Fundamental Rights; art. 28 of the Data Protection Directive.

Apart from the general Data Protection Directive, the e-Privacy Directive is also relevant for behavioural targeting.[25] Since 2009, article 5(3) of the e-Privacy Directive requires any party that stores or accesses information on a user's device to obtain the user's informed consent. Article 5(3) applies regardless of whether personal data are processed, and applies to many tracking technologies such as tracking cookies. There are exceptions to the consent requirement, for example for cookies that are necessary for communication or for a service requested by the user. Hence, no prior consent is needed for cookies that are used for log-in procedures, or for a digital shopping cart. There is much discussion on how consent for cookies should be obtained.[26] Many marketers suggest that people give implied consent if they do not object to tracking cookies.[27] The Article 29 Working Party, however, says and expression of will is required for valid consent.[28]

Article 5(3) of the e-Privacy Directive could be seen as a sector-specific rule for tracking through cookies and similar technologies, which complements the general data protection regime. But article 5(3) does not say anything about the data that are collected with cookies. Hence, article 5(3) does not regulate storing, analysing, combining, selling, or securing data that are collected for behavioural targeting. But, as far as personal data are processed, data protection law does regulate such activities. From here on, this paper focuses on the general Data Protection Directive, rather than on the e-Privacy Directive or the national laws implementing the directives.

---

[25] See Eleni Kosta, 'Peeking into the Cookie Jar: The European Approach towards the Regulation of Cookies', International Journal of Law and Information Technology (2013) 17.

[26] See: Ronald Leenes and Eleni Kosta, 'Taming the cookie monster with Dutch law–A tale of regulatory failure' [2015] 31 Computer Law & Security Review, 317.

[27] InteractiveAdvertisingBureauUnitedKingdom,'DepartmentforBusiness, Innovation & Skills Consultation on Implementing the Revised EU Electronic Communications Framework, IAB UK Response' (1 December 2012) <www.iabuk.net/sites/default/files/IABUKresponsetoBISconsultationonimplementingt herevisedEUElectronicCommunicationsFramework_ 7427_0.pdf> accessed 24 January 2016, p. 2.

[28] See, eg, Article 29 Working Party, 'Opinion 15/2011 on the definition of consent' (WP 187) 13 July 2011, pp. 32, 35.

## 4    Behavioural targeting data are used to single out people

Data protection law only applies if 'personal data' are processed. Almost everything that can be done with personal data, such as collecting, storing, or analysing data, falls within the definition of 'processing'.[29] But do companies process 'personal data' when they use nameless individual profiles for behavioural targeting? The personal data definition in the Data Protection Directive reads as follows:

> 'Personal data' shall mean any information relating to an identified or identifiable natural person ('data subject'); an identifiable person is one who can be identified, directly or indirectly, in particular by reference to an identification number or to one or more factors specific to his physical, physiological, mental, economic, cultural or social identity.[30]

Personal data are thus not limited to a name and address, but include all information that relates to a person who can be identified, directly or indirectly. The Court of Justice of the European Union confirms that information without a name can constitute personal data.[31]

European Data Protection Authorities, cooperating in the Article 29 Working Party, say that behavioural targeting generally entails personal data processing, because companies use the data to *single out* individuals. The Working Party is an independent advisory body and publishes opinions on the interpretation of data protection law.[32] Although not legally binding, the opinions are influential. Judges and Data Protection Authorities often follow the Working Party's interpretation.[33]

---

[29] Art. 2(b) of the Data Protection Directive defines personal data processing. See on the directive's scope: Art. 1(1).

[30] Ibid., Art. 2(a), capitalisation adapted.

[31] Case C-101/01, *Lindqvist,* EU:C:2003:596, par 27: 'identifying [people] by name *or by other means,* for instance by giving their telephone number or information regarding their working conditions and hobbies, constitutes 'the processing of personal data (…)'. See also: Case C-92/09 and C-93/09, *Volker und Markus Schecke and Eifert,* EU:C:2010:662; Case C-468/10 and C 469/10, ASNEF, EU:C:2011:777, par. 27.

[32] See Art. 29 of the Data Protection Directive.

In a 2007 opinion on the concept of personal data, the Working Party notes that four elements in the personal data definition provided in the Data Protection Directive can be distinguished: (i) 'any information', (ii) 'relating to', (iii) 'an identified or identifiable', (iv) 'natural person'.[34] I will analyse behavioural targeting in the light of these four elements.

*Element 1: 'any information'* – Data collected for behavioural targeting, such as data about people's web browsing history, fall within the scope of any information.[35]

*Element 2: 'relating to'* – Sometimes information relates to a person because it refers to an object, such as a computer or a car. The Court of Justice of the European Union confirms that data relating to an object can identify a person.[36] A behavioural targeting company often recognises a person's device, such as a computer or a smart phone.

Information can relate to a person because of its content, its purpose, or its result, according to the Working Party. Information relates to a person because of its *content* when it is 'about' a person.[37] For instance, information in a patient's medical file clearly relates to the patient, regardless of the purpose or the result of using this information.[38] If a company holds a nameless individual profile for behavioural targeting, the information in the profile will relate to a person because of its *content*. For example, the company might possess a list of websites visited and a list of inferred interests for the person with a cookie with ID *xyz* on his or her computer.

---

[33] See: Serge Gutwirth and Yves Poullet, 'The contribution of the Article 29 Working Party to the construction of a harmonised European data protection system: an illustration of 'reflexive governance'?', in Pablo Palazzi, María Verónica Pérez Asinari (eds), *Défis du Droit à la Protection de la Vie Privée. Challenges of Privacy and Data Protection Law* (Bruylant 2008).
[34] See Article 29 Working Party, 'Opinion 4/2007 on the concept of personal data' (WP 136), 20 June 2007.
[35] *Ibid.*, pp. 6-9.
[36] In Lindqvist, the ECJ mentions a phone number as an example of information that can identify somebody, while a phone number relates to an object (a phone) rather than to a person (Case C-101/01, *Lindqvist*, EU:C:2003:596, par 27). See also section 5 below, on IP addresses in Case C-70/10, *Scarlet v Sabam*, EU:C:2011:771.
[37] Article 29 Working Party 2007, WP 136, *Op.cit.*, pp. 9-10.
[38] Ibid., p. 10.

Behavioural targeting information may also relate to a person because of its *result*. If a company targets an ad to a specific person based on information about this person, the company treats this person differently from others: a result. Information can also relate to a person because of its *purpose*, if a company uses data 'with the <u>purpose</u> to evaluate, treat in a certain way or influence the status or behaviour of an individual'.[39] A behavioural targeting company processes data about an individual to influence this individual, to make him or her click on an ad. The Working Party concludes that information processed for behavioural targeting '*relates to, (i.e.* is about) a person's characteristics or behaviour and it is used to influence that particular person.'[40]

In a 2014 case, the Court of Justice of the European says that a 'legal analysis' in a dossier about an asylum seeker is not, in itself, a piece of personal data.[41] Arguably, however, such a legal analysis relates to the asylum seeker because of the result: will asylum be granted or not? This judgment does probably not imply that data that relate to a person because of their result should generally not be seen as personal data. The judgment concerns a specific situation: access to dossiers in asylum procedures. Moreover, the judgment does not explicitly reject the Working Party's view on data that relate to a person because of a result.[42]

Some data processing activities for behavioural targeting do not concern personal data. A company can use personal data to construct a model, along the following lines: *0,2% of people who visit websites about cycling clicks on ads for bikes, while 0.1% of random people clicks on such ads.* Many 'big data' practices involve similar models.[43] Such models do not consist of personal data, as they do not relate to a specific person.[44]

---

[39] Ibid., p. 10 (emphasis original).
[40] The Working Party concludes that information processed for behavioural targeting '*relates to, (i.e.* is about) a person's characteristics or behaviour and it is used to influence that particular person.' Article 29 Working Party, 'Opinion 2/2010 on online behavioural advertising' (WP 171), 22 June 2010, p. 9 (emphasis original).
[41] Case C-141/12 and C-372/12, *YS. and M. and S*, ECLI:EU:C:2014:2081.
[42] See: Evelien Brouwer and Frederik Zuiderveen Borgesius, 'Access to Personal Data and the Right to Good Governance during Asylum Procedures after the CJEU's *YS. and M. and S*. judgment (C-141/12 and C-372/12)', European Journal of Migration and Law, 2015-7, p. 259-272; Mark Jansen, 'Arrest HvJ EU inzake begrip persoonsgegevens en karakter inzagerecht', *Privacy and Informatie* (2014/5), 200–206.

As soon as a company applies the model to an individual, however, the information relates to this person because of its purpose or its result. For instance, if somebody who has a cookie with the ID *xyz* on his computer visits a website, an ad network may recognise this person (the cookie) as somebody who visits many websites about cycling. According to the company's model, people who visit websites about cycling are likely to click on ads for bikes. Therefore, the company shows the person bike advertising. At that moment, the company singles out this individual, applies the model to this individual, and aims to influence this individual. [45] In sum, behavioural targeting usually entails processing 'information relating to' a person.[46]

*Element 3: 'an identified or identifiable'* – Does behavioural targeting entail the processing of data that relates to an 'identifiable' person? In other words, does a behavioural targeting company process data by which a person can be 'directly or indirectly identified'?

The personal data definition refers to an 'identification number' as an example of information by which a person can be identified. When the Data Protection Directive was drafted in the early 1990s, the EU lawmaker probably thought of social security numbers and similar identifiers.

---

[43] See Eric Siegel, *Predictive Analytics: The Power to Predict Who Will Click, Buy, Lie, or Die* (John Wiley & Sons 2013), p. 26.

[44] See Wim Schreurs, Mireille Hildebrandt, Els Kindt, Michaël Vanfleteren, 'Cogitas, Ergo Sum. The Role of Data Protection Law and Non-discrimination Law in Group Profiling in the Private Sector', in Mireille Hildebrandt and Serge Gutwirth (eds), *Profiling the European citizen: Cross-Disciplinary Perspectives* (Springer 2008). See also B. Custers, The Power of Knowledge, Ethical, Legal, and Technological Aspects of Data Mining and Group Profiling in Epidemiology, Nijmegen: Wolf Legal Publishers 2004, p. 151.

[45] See Bert Jaap Koops, Some reflections on profiling, power shifts, and protection paradigms, in Mireille Hildebrandt and Serge Gutwirth (eds), *Profiling the European citizen: cross-disciplinary perspectives* (Springer 2008), p. 331; See also B. Custers, The Power of Knowledge, Ethical, Legal, and Technological Aspects of Data Mining and Group Profiling in Epidemiology, Nijmegen: Wolf Legal Publishers 2004, p. 151.

[46] Art. 2(a) of the Data Protection Directive.

Nevertheless, unique identifiers in cookies are strings of numbers and letters, and can thus be seen as identification numbers.[47]

Furthermore, a cookie with a unique identifier enables a company to follow somebody's online behaviour to infer his or her interests. In a 2007 opinion, the Working Party stated that persons are identifiable if they can be distinguished within a group: 'singling out' an individual implies identifying this individual.[48] 'In fact, to argue that individuals are not identifiable, where the purpose of the processing is precisely to identify them, would be a sheer contradiction in terms.'[49] In later opinions, the Working Party explicitly says that cookies and similar files with unique identifiers are personal data, because 'such unique identifiers enable data subjects to be 'singled out' for the purpose of tracking user behaviour while browsing on different websites and thus qualify as personal data.'[50]

However, it is possible to think of scenarios in which behavioural targeting data tied to a unique identifier are not personal data, because the data do not relate to an individual. For instance, a computer in an Internet café might be used by many people.[51] An ad network which sets a cookie on this computer might compile a profile based on the surfing behaviour of many people. Such a profile should probably not be considered personal data. Nevertheless, if a company uses a unique identifier for behavioural targeting, the identifier usually relates to one person. Furthermore, some companies can recognise people on the basis of their browsing behaviour, because any person's browsing behaviour is unique.[52]

---

[47] See Colette Cuijpers, Arnold Roosendaal and Bert Jaap Koops, D11.5: The legal framework for location-based services in Europe (Future of Identity in the Information Society, FIDIS, 12 June 2007) <www.fidis.net/fileadmin/fidis/deliverables/fidis-WP11-del11.5-legal_framework_for_LBS.pdf> accessed 24 January 2016, p. 25. See also Peter Traung, 'EU Law on Spyware, Web Bugs, Cookies, etc. Revisited: Article 5 of the Directive on Privacy and Electronic Communications' (2010) 31 Business Law Review 216.
[48] Article 29 Working Party 2007, WP 136, *Op.cit.*, p. 12-14.
[49] Ibid., p. 16. The Working Party does not make this remark in the context of behavioural targeting.
[50] Article 29 Working Party, 'Opinion 16/2011 on EASA/IAB Best Practice Recommendation on Online Behavioural Advertising' (WP 188) 8 December 2011, p. 8. See also Article 29 Working Party, 'Opinion 1/2008 on data protection issues related to search engines' (WP 148), 4 April 2008.
[51] The example is taken from Article 29 Working Party 2007, WP 136, *Op.cit.*, p. 17.

*Element 4: 'natural person'* – A natural person is not a legal person,[53] and is not a deceased person.[54] With behavioural targeting, companies usually process information about natural persons.

The Working Party is not alone in its interpretation of the personal data definition. For instance, the UK Information Commissioner's Office says that behavioural targeting entails the processing of personal data.[55] A Dutch statute even contains a legal presumption: using tracking cookies for behavioural targeting is presumed to entail the processing of personal data.[56] In an investigation of behavioural targeting company YD (now called Yieldr), the Dutch Data Protection Authority confirms that a name is not necessary to identify a person, and concludes that YD processes personal data.[57] The Authority arrives at a similar conclusion in an investigation regarding tracking people's viewing behaviour on smart TVs.[58]

---

[52] See e.g. Claire Cain Miller and Somini Sengupta, 'Selling Secrets of Phone Users to Advertisers', New York Times 5 October 2013, <www.nytimes.com/2013/10/06/technology/selling-secrets-of-phone-users-to-advertisers.html> accessed 24 January 2016.

[53] This is a bit more complicated. For instance, data about a sole proprietorship may be personal data. See Bart van der Sloot, 'Do privacy and data protection rules apply to legal persons and should they? A proposal for a two-tiered system' [2015] 31 Computer Law & Security Review (2015) 26. See also Bygrave LA, *Data protection law: approaching its rationale, logic and limits (PhD thesis University of Oslo),* vol 10 (Information Law Series, Kluwer Law International 2002), chapter 9-16.

[54] See on the question of whether privacy rights do – or should – continue after death: the special issue of SCRIPTed: (2013) 10(1). See also David Korteweg and Frederik Zuiderveen Borgesius, 'E-mail na de dood. Juridische bescherming van privacybelangen' [Post-mortem email. legal protection of privacy interests] (2009)(5) Privacy & Informatie.

[55] Information Commissioner, 'Personal Information Online. Code of Practice' (2010) <https://ico.org.uk/media/for-organisations/documents/1591/personal_information_online_cop.pdf> accessed 24 January 2016, p. 22: 'The end result of this [behavioural targeting] process is that particular content is displayed to particular individuals, depending on the 'score' that is associated with their inferred interests. Allocating the score and using it to target advertising involves the processing of personal data and the DPA applies.'

[56] Art. 11.7a of the Dutch Telecommunications Act (See for a translation Frederik J. Zuiderveen Borgesius, Behavioral Targeting. Legal Developments in Europe and the Netherlands' (position paper for W3C Workshop: Do Not Track and Beyond 2012) <www.w3.org/2012/dnt-ws/position-papers/24.pdf> accessed 24 January 2016).

In a letter to Google, signed by 27 national Data Protection Authorities, the Working Party says that Google processes personal data about its 'passive users.'[59] These are people that are tracked through Google's ad network DoubleClick.[60] Along similar lines, the French Data Protection Authority CNIL says about Google:

> [T]he sole objective pursued by the company is to gather a maximum of details about individualized persons in an effort to boost the value of their profiles for advertising purposes. Its business model then is not dependent on knowing the last name, first name, address or other directly identifying details about individuals, which it does not need to recognize them every time they use its services. (…) In other words, the accumulation of data that it holds about any one person allows it to individualize the person based on one or more uniquely personal details. These data

---

[57] College bescherming persoonsgegevens, 'Onderzoek naar de verwerking van persoonsgegevens door YD voor behavioural targeting. Rapport definitieve bevindingen Maart 2014 met corrigendum van 29 april 2014' [Investigation personal data processing by YD for behavioural targeting. Public version Report definitive findings with correction of 29 April 2014] (Z2012-00811)] (13 May 2014) <www.cbpweb.nl/sites/default/files/atoms/files/rap_2013_yd-cookies-privacy.pdf> accessed 24 January 2016, p. 72.

[58] College bescherming persoonsgegevens, 'Onderzoek naar de verwerking van persoonsgegevens met of door een Philips smart tv door TP Vision Netherlands B.V. Openbare versie Rapport definitieve bevindingen' [Investigation perrsonal data processing through Philips smart TV by TP Vision Netherlands BV. Public version report definitive findings] (z2012-00605)' (July 2013) <https://cbpweb.nl/sites/default/files/downloads/pb/pb_20130822-persoonsgegevens-smart-tv.pdf> accessed 17 February 2014, p. 59.

[59] Article 29 Working Party, Letter to Google (signed by 27 national Data Protection Authorities), 16 October 2012 <www.cnil.fr/fileadmin/documents/en/20121016-letter_google-article_29-FINAL.pdf> Appendix: <www.cnil.fr/fileadmin/documents/en/GOOGLE_PRIVACY_POLICY-_RECOMMENDATIONS-FINAL-EN.pdf> accessed 24 January 2016.

[60] Ibid., appendix, p. 2, footnote 2. Passive users are 'users who does not directly request a Google service but from whom data is still collected, typically through third party ad platforms, analytics or +1 buttons.'

must, as such, be considered as identifiable rather than anonymous.[41]

Outside Europe, regulators have diverging opinions. According to the Australian Privacy Commissioner for example, behavioural targeting data does not necessarily identify individuals.[42] Other regulators hold views similar to European Data Protection Authorities. The International Working Group on Data Protection in Telecommunications, with members from around the world, says that behavioural targeting usually entails the processing of personal data.[43] Similarly, the Privacy Commissioner of Canada says that behavioural targeting generally entails personal data processing.[44] The personal data definition in a 2014 report by the US Federal Trade Commission includes 'a persistent identifier, such as a customer number held in a 'cookie' or processor serial number'.[45] In sum, several

---

[41] Commission Nationale de l'Informatique et des Libertés, 'Deliberation No. 2013-420 of the Sanctions Committee of CNIL imposing a financial penalty against Google Inc' (8 January 2014, English translation) (8 January 2014) <www.cnil.fr/fileadmin/documents/en/D2013-420_Google_Inc_EN.pdf> accessed 24 January 2016, p. 11. The Dutch Data Protection Authority confirms this interpretation in a report on Google and DoubleClick. College bescherming persoonsgegevens 2013 (Google) – 'Investigation into the combining of personal data by Google, Report of Definitive Findings' (z2013-00194) [English translation of Onderzoek CBP naar het combineren van persoonsgegevens door Google, Rapport definitieve bevindingen (z2013-00194)] (November 2013, with correction 25 November 2013) <www.cbpweb.nl/sites/default/files/downloads/mijn_privacy/en_rap_2013-google-privacypolicy.pdf> accessed 24 January 2016, p. 44: 'Identification is also possible without finding out the name of the data subject. All that is required is that the data can be used to distinguish one particular person from others.' See also p. 49-57.

[42] 'The information collected by online advertisers may often not be sufficient to identify you; it might just be general information about your interests and sites you have visited.' Office of the Australian Privacy Commissioner, 'Privacy Fact Sheet 4: Online Behavioural Advertising Know Your Options' (December 2011) <www.oaic.gov.au/privacy/privacy-resources/privacy-fact-sheets/other/privacy-fact-sheet-4-online-behavioural-advertising-know-your-options> accessed 24 January 2016.

[43] International Working Group on Data Protection in Telecommunications, Web Tracking and Privacy (July 2013) <www.datenschutz-berlin.de/attachments/949/675.46.13.pdf> accessed 24 January 2016.

[44] Office of the Privacy Commissioner of Canada, 'Privacy and Online Behavioural Advertising' (Guidelines, June 2012) <www.priv.gc.ca/information/guide/2011/gl_ba_1112_e.pdf> accessed 24 January 2016, p. 2.

[45] Federal Trade Commission, 'Data Brokers. A Call for Transparency and Accountability' (May 2014) <www.ftc.gov/system/files/documents/reports/data-brokers-call-transparency-accountability-report-federal-trade-commission-may-2014/140527databrokerreport.pdf> accessed 24 January 2016, Appendix A, p. A16.

regulators outside Europe agree with the Working Party's view on identifiability.

In conclusion, singling out a person implies identifying this person, even if the data controller cannot tie a name to the data it processes about an individual. Therefore, behavioural targeting generally entails personal data processing. Moreover, regardless of whether singling out an individual implies identifying this individual, it is often fairly easy to tie a name to behavioural targeting data, as discussed in the next section.

## 5    Behavioural targeting data can be tied to people's names

The company holding nameless data, or another party, can often add a name to the data, and thus identify the data subject by name. To structure the analysis, I distinguish between four situations in which a company processes data about a person.

(i)    A company processes data about an individual, and it knows the individual's name.

(ii)    A company processes data about an individual, and it is fairly easy for the company to tie a name to the data.

(iii)    A company processes data about an individual, and it is difficult for the company to add a name to the data, but it would be fairly easy for another party to tie a name to the data.

(iv)    A company processes data about an individual, and it would be difficult for anybody to tie a name to the data.

For situations (i) and (ii) it is not controversial that data protection law applies. In situation (i), a person is identified; in situation (ii), a person is clearly identifiable. Situations (iii) and (iv) provoke more discussion.

*Situation (i) – A company knows the person's name*

In situation (i), a behavioural targeting company processes data about an individual, and it knows the individual's name. To illustrate: a provider of a social network site often holds individual profiles with names. For instance, Facebook, which tracks its users around the web for targeted marketing,[46] requires its users to register under their own name.[47] Such a company processes data about an identified person. Google also tracks people around the web for behavioural targeting and may know the name of people who have a Google account.[48]

*Situation (ii) – A company can add the person's name*

In situation (ii), a company processes data about an individual, and it is fairly easy for the company to tie a name to the data. The preamble of the Data Protection Directive says: 'To determine whether a person is identifiable, account should be taken of *all the means likely reasonably to be used* either by the controller or by any other person to identify the said person.'[49]

The question is thus: which means can a company that processes data about a person 'reasonably likely use' to identify a person?[50] The answer depends,

---

[46] Güneş Acar, Brendan Van Alsenoy, Frank Piessens, Claudia Diaz, Bart Preneel, 'Facebook Tracking Through Social Plug-ins' (Technical report prepared for the Belgian Privacy Commission) (V. 1.1, 24 June 2015) <https://securehomes.esat.kuleuven.be/~gacar/fb_tracking/fb_plugins.pdf> accessed 24 January 2016.

[47] Facebook's Name Policy <www.facebook.com/help/292517374180078> accessed 24 January 2016.

[48] See Article 29 Working Party, Letter to Google (signed by 27 national Data Protection Authorities, 16 October 2012 <www.cnil.fr/fileadmin/documents/en/20121016-letter_google-article_29-FINAL.pdf> Appendix: <www.cnil.fr/fileadmin/documents/en/GOOGLE_PRIVACY_POLICY-_RECOMMENDATIONS-FINAL-EN.pdf> accessed 24 January 2016.

[49] Recital 26 of the Data Protection Directive (emphasis added).

[50] Following the definition of 'data subject' (Art. 4(1)) of the European Commission proposal for a Data Protection Regulation), this paper switches the words 'likely reasonably' to

among other things, on the state of science and technology, and on how costly it would be to identify somebody. According to the Working Party, 'a mere hypothetical possibility to single out the individual is not enough to consider the person as 'identifiable.''[71]

It is often possible to identify people within a purportedly anonymised data set. In 2000, Sweeney found that 87% of the US population is uniquely identified by three attributes: their date of birth, their gender, and their ZIP code. [72] Techniques to re-identify data subjects continue to improve. [73] Furthermore, purportedly anonymised data may be re-identified when they are combined with other data.[74] More and more data sets become available, for instance from social network sites, which can be merged with the original data set. Computer scientists summarise that de-identifying personal data is an 'unattainable goal.' [75] For instance, people can be identified on the basis of location data. Research has shown that 'in a dataset where the location of an individual is specified hourly, and with a spatial resolution equal to that given by the carrier's antennas, four spatio-temporal points are enough to uniquely identify 95% of the individuals.[76]

---

'reasonably likely'. (European Commission, Proposal for a Regulation of the European Parliament and of the Council on the Protection of Individuals with regard to the Processing of Personal Data and on the Free Movement of Such Data (General Data Protection Regulation) COM(2012) 11 final, 2012/0011 (COD), 25 January 2012).

[71] Article 29 Working Party 2007, WP 136, *Op.cit.*, p. 15. See also: Article 29 Working Party, 'Opinion 05/2014 on Anonymisation Techniques' (WP 216) 10 Apr. 2014, pp. 8-9.

[72] Latanya Sweeney, Simple Demographics Often Identify People Uniquely, Data Privacy Working Paper 3 Pittsburgh 2000 <http://dataprivacylab.org/projects/identifiability/paper1.pdf>. See also Latanya Sweeney and Ji Su Yoo, 'De-anonymizing South Korean Resident Registration Numbers Shared in Prescription Data', Technology Science. 2015092901, 29 September 2015, <http://techscience.org/a/2015092901>; Latanya Sweeney, 'Only You, Your Doctor, and Many Others May Know', Technology Science 2015092903, 29 September 2015 <http://techscience.org/a/2015092903>, all accessed 24 January 2016.

[73] Matthijs Koot, 'Measuring and Predicting Anonymity' (PhD thesis University of Amsterdam) (2012) <https://cyberwar.nl/d/PhD-thesis_Measuring-and-Predicting-Anonymity_2012.pdf> accessed 24 January 2016.

[74] Arvind Narayanan, Joanna Huey and Edward W. Felten. 'A Precautionary Approach to Big Data Privacy' (19 March 2015) <http://randomwalker.info/publications/precautionary.pdf> accessed 24 January 2016.

[75] Arvind Narayanan and Vitaly Shmatikov, Myths and Fallacies of Personally Identifiable Information (2010) 53(6) Communications of the ACM 24-26, p. 26.

[76] Yves-Alexandre de Montjoye, César A. Hidalgo, Michel Verleysen, Vincent D Blondel, 'Unique in the Crowd: The privacy bounds of human mobility' (2013) 3 Nature, Scientific reports.

Sometimes, the person behind nameless data can be found without any sophisticated data analysis. In 2006, search engine provider AOL released a data set of nameless search profiles, each tied to a random number. Within a few days, *New York Times* journalists had found one of the searchers: 'A face is exposed for AOL searcher no. 4.417.749.' [77] The search queries suggested that the searcher was an elderly woman with a dog, from the town of Lilburn. An interview confirmed that the journalists had correctly identified her.

A behavioural targeting company can often tie a name to the data it holds about an individual, considering 'the means reasonably likely to be used' by the company. For instance, some companies offer services to consumers directly. If a company has a cookie-based profile of a person, and offers an email service to the same person, the company can tie the person's email address to the cookie. Most email addresses are personal data. [78] In addition, email addresses and email messages often contain the user's name. If a company offers a social network site or another service where people log in, the company can also tie the log-in information (such as a user names) to its cookies.

A search engine company that holds nameless user profiles with search queries might be able to identify people based on their searches. As a Google employee said in a court case, 'there are ways in which a search query alone may reveal personally identifying information.' [79] If the user sometimes searches for his or her name, it would be even easier to attach a name to the user profile. [80] In sum, a company that processes nameless data about individuals can often fairly easily tie a name to the data: the company processes personal data.

---

[77] Michael Barbarom and Tom Zeller, 'A Face Is Exposed for AOL Searcher No. 4417749' (New York Times, 9 August 2006) <www.nytimes.com/2006/08/09/technology/09aol.html> accessed 24 January 2016.
[78] An 'info@' email address of a company might not constitute personal data, if it does not refer to an individual.
[79] Matt Cutts, Declaration in Gonzales v. Google, 234 F.R.D. 674 (N.D. Cal. 2006) (17 February 2006) <http://docs.justia.com/cases/federal/district-courts/california/candce/5:2006mc80006/175448/14/0.pdf> accessed 24 January 2016, p. 9.
[80] See on such vanity searchers: Chris Soghoian, 'The Problem of Anonymous Vanity Searches' I/S: A Journal of Law & Policy for the Information Society (2007) 3 299.

***Situation (iii) – Another party can add the person's name***

In situation (iii), a behavioural targeting company processes data about an individual, and it is difficult for the company to add a name to the data, but it would be fairly easy for *another party* to do so. Whereas it is commonly accepted that companies in situations (i) and (ii) process personal data, situation (iii) still leads to discussion.

To give a hypothetical example of situation (iii): an ad network has a cookie-based profile of a person, including an IP address. Let's assume, for the sake of argument, that it would be difficult for the ad network to tie a name to the profile and the IP address. However, the person's Internet access provider can tie a name to the IP address. For an online shop this would be easy too, if the person orders a product and provides the shop with his or her name and address.

Does it matter that only another party can attach a name to the data? Recital 26 of the Data Protection Directive suggests the answer is No: 'To determine whether a person is identifiable, account should be taken of all the means reasonably likely to be used either by the controller *or by any other person* to identify the said person.'[81] The approach adopted in the Recital is sometimes called the absolute approach.[82] A relative approach would imply only looking at the means that the data controller (the company) is likely to use.[83] In 2007, the Working Party suggested that, in the context of medical research, a relevant approach to identifiability could be taken.[84] But in a 2014 opinion, the Working Party clearly chose the absolute approach.[85]

---

[81] Emphasis added.
[82] See: European Commission's Information Society and Media Directorate-General, Legal analysis of a Single Market for the Information Society, chapter 4: The future of online privacy and data protection, prepared by DLA Piper 2011 <http://ec.europa.eu/information_society/newsroom/cf/itemdetail.cfm?item_id=7022> accessed 24 January 2016, pp. 18-21.
[83] See for the definition of the data controller: Art. 2(d) of the Data Protection Directive. As noted, the distinction between data controllers and data processors falls outside the scope of this paper.

While Recital 26 suggests an absolute approach, the means at the disposal of the data controller are relevant to determine which means are reasonably likely to be used for identification. For instance, if a random person finds some human hairs, those hairs are not personal data for the finder. But if the police found those hairs and can match them against a DNA database, the hairs should probably be regarded as personal data.[86]

Recitals do not have the same legal weight as the provisions of a directive. Nevertheless, when interpreting the provisions of a directive, the Court of Justice of the European Union often considers the recitals.[87] The Court also refers to recitals in data protection cases.[88] Klimas & Vaiciukaite summarise: 'Recitals in EC law are not considered to have independent legal value, but they can expand an ambiguous provision's scope.'[89] Presumably, Recital 26 can be used to interpret the personal data definition.

Sometimes, Data Protection Authorities say that personal data are identifiable for one party, while they are not identifiable for another party – a relative approach. [90] Hence, Data Protection Authorities sometimes consider which means the company holding the data can use. For example, the English Information Commissioner's Office appears to favour the relative approach. [91] In sum, while Recital 26 seems to dictate an absolute approach, the relative approach may be relevant when determining which means are likely to be used for identification.

---

[84] Article 29 Working Party 2007, WP 136, *Op.cit.*, p 19. See also question 7 on this website about the US data protection safe harbour <http://www.export.gov/safeharbor/eu/eg_main_018386.asp> accessed 24 January 2016.

[85] Article 29 Working Party 2014, WP 216, *Op.cit.*, p. 9: 'it is critical to understand that when a data controller does not delete the original (identifiable) data at event-level, and the data controller hands over part of this dataset (for example after removal or masking of identifiable data), the resulting dataset is still personal data.'

[86] The example is inspired by an example by G.J. Zwenne, '*De verwaterde privacywet' [Diluted Privacy Law], Inaugural lecture of Prof. Dr. G. J. Zwenne to the office of Professor of Law and the Information Society at the University of Leiden on Friday,* 12 April 2013 <http://zwenneblog.weblog.leidenuniv.nl/files/2013/09/G-J.-Zwenne-De-verwaterde-privacywet-oratie-Leiden-12-apri-2013-NED.pdf> accessed 24 January 2016.

[87] General Secretariat Of The Council Of The European Union, 'Manual Of Precedents For Acts Established Within The Council Of The European Union', 9 July 2010, <http://ec.europa.eu/translation/maltese/guidelines/documents/form_acts_en.pdf> accessed 24 January 2016.

[88] See e.g. Case C-131/12, *Google Spain*, EU:C:2013:424, par. 48, 54, 58, 66-67; Case C-101/01, *Lindqvist*, EU:C:2003:596, par. 95.

[89] Tadas Klimas and Jūratė Vaičiukaitė, 'The Law of Recitals in European Community Legislation' ILSA Journal of International & Comparative Law (2008)(15), p. 2-33, p. 3. 'Where the recital is clear, it will control an ambiguous operative provision. This means that the operative provision will be interpreted in the light of the recital' (p. 33).

There are many ways for behavioural targeting companies to attach a name to data. Computer scientist Narayanan discusses some examples. For instance, many websites disclose identifying information about their visitors to ad networks, often inadvertently. Furthermore, some companies specialise in tying names to data held by ad networks. The goal of some web surveys – 'Win a free smart phone!' – is to match email addresses and names to behavioural targeting data. If you provide your email address to a company that also operates a cookie, the company can tie the two together. If one company tied a name to a cookie profile, it can provide the name to other companies that only had a nameless profile, through cookie synching. Narayanan summarises that 'there is no such thing as anonymous online tracking.' [92]

The discussion about behavioural targeting data resembles the discussion about IP addresses. The Working Party and many judges in Europe say that IP addresses generally qualify as personal data.[93] Others disagree. First, some argue for a relative approach. For instance, Google suggests that an IP address should not be seen as personal data if the company holding the IP address cannot tie a name to it.[94] Second, IP addresses sometimes cannot be used to identify a person.[95] For example, some organisations, such as the University of Amsterdam (my workplace), access the internet through one IP address. And the country Qatar routed all internet traffic through a couple of IP addresses.[96] In such cases, a mere IP address without any other information may not be enough to identify somebody.

---

[90] Impact Assessment for the proposal for a Data Protection Regulation (SEC(2012) 72 final of 25 Jan. 2012), Annex 2, pp. 15-16.
[91] Information Commissioner's Office 2012 – Anonymisation: Managing Data Protection Risk Code of Practice (November 2012) <http://ico.org.uk/for_organisations/data_protection/topic_guides/~/media/documents/library/Data_Protection/Practical_application/anonymisation-codev2.pdf> accessed 24 January 2016, p. 21. See also: Leslie Stevens, 'The Proposed Data Protection Regulation and Its Potential Impact on Social Sciences Research in the UK' (2015)(2) European Data Protection Law Review 97. The German situation is more complicated, but also boils down to a relative approach (see Douwe Korff, 'New Challenges to Data Protection Study Country Report: Germany' (European Commission DG Justice, Freedom and Security Report 2010) <http://ec.europa.eu/justice/policies/privacy/docs/studies/new_privacy_challenges/final_report_country_report_A4_germany.pdf> accessed 24 January 2016, p. 4).
[92] Arvind Narayanan, 'There is no such thing as anonymous online tracking' (Center for Internet and Society, Stanford Law School, 28 July 2011) <https://cyberlaw.stanford.edu/blog/2011/07/there-no-such-thing-anonymous-online-tracking> accessed 24 January 2016.

In the 2012 Scarlet/Sabam case, the Court of Justice of the European Union decided that the IP addresses in the case were personal data. Copyright organisation Sabam requested Internet access provider Scarlet to install a filtering system to enforce copyrights. Scarlet refused. The Court decided that the IP addresses are personal data: 'Those addresses are protected personal data because they allow those users to be precisely identified.'[97] The Advocate General referred to opinions by the Working Party to support his conclusion that the IP addresses were personal data.[98]

Still, the discussion about IP addresses is not over. Although it used ambiguous language in Scarlet/Sabam, the Court may have suggested a relative approach.[99] For companies that are not Internet access providers, it is harder to tie an IP address to a name. Such companies might argue that IP addresses are not personal data in their hands. In his 2013 opinion for the Google Spain case, the Advocate General viewed IP addresses as personal data when they are in the hands of Google.[100] This suggests an absolute approach to identifiability. The Court of Justice of the European Union neither confirmed nor disproved this view in the subsequent judgment. Soon we may have more certainty: the German Bundesgerichtshof has asked the Court of Justice of the European Union whether dynamic IP addresses constitute personal data when they are not in the hands of an access provider.[101]

---

[93] See about the status of IP addresses as personal data: Impact Assessment for the proposal for a Data Protection Regulation (SEC(2012) 72 final of 25 January 2012), Annex 2, pp. 14-16; Time.lex, 'Study of case law on the circumstances in which IP addresses are considered personal data' SMART 2010/12 D3. Final report (May 2011) <www.timelex.eu/frontend/files/userfiles/files/publications/2011/IP_addresses_report_-_Final.pdf> accessed 24 January 2016; Article 29 Working Party 2007, WP 136, *Op.cit*, p. 17.
[94] See e.g. Alma Whitten, Are IP addresses personal? (Google Public Policy Blog, 22 Feb. 2008) <http://googlepublicpolicy.blogspot.com/2008/02/are-ip-addresses-personal.html> accessed 24 January 2016. See also Zwenne 2013, *Op.cit*, p. 26.
[95] Zwenne 2013, *Op.cit*, p. 28.
[96] Jonathan Zittrain, *The Future of the Internet and How to Stop It* (Yale University Press 2008), p. 157.
[97] Case C-70/10, *Scarlet v Sabam*, EU:C:2011:771, par 51.
[98] Opinion AG (14 April 2011) for Case C-70/10, *Scarlet v Sabam*, EU:C:2011:255, par. 75-80.
[99] In an earlier publication I assumed that the ECJ limited its remark to IP addresses in the hands of access provider Scarlet (Stefan Kulk and Frederik Zuiderveen Borgesius, Filtering for Copyright Enforcement in Europe after the Sabam Cases (2012) 34(11) European Intellectual Property Review 791-795). Now I believe the CJEU may have taken an absolute approach, as the CJEU talks about 'users', and not about 'subscribers'.

The case law on IP addresses is relevant for the discussion about behavioural targeting profiles, because it confirms that nameless data referring to a device can be personal data. Furthermore, companies that use behavioural targeting usually tie IP addresses to the data they process. For instance, an ad network typically needs the IP address of a device to deliver an ad.

But there is an important difference between IP addresses and individual profiles that are used for behavioural targeting. Behavioural targeting profiles usually contain much more information than an IP address.[102] Hence, adding a name to a behavioural targeting profile is often easier than adding a name to an IP address.

To conclude, if a company processes nameless data about an individual, and it is fairly easy for another party to tie a name to the data, it follows from the Data Protection Directive's preamble that the data are personal data.

---

[100] Opinion AG, Case C-131/12, *Google Spain*, EU:C:2013:424, par. 3 and par. 48: 'An internet search engine service provider may automatically acquire personal data relating to its users (...). This automatically transmitted data can include their IP address, user preferences (language, etc.)'

[101] Case C-582/14, *Breyer*: 'Must Article 2(a) of (…) the Data Protection Directive (…) be interpreted as meaning that an Internet Protocol address (IP address) which a service provider stores when his website is accessed already constitutes personal data for the service provider if a third party (an access provider) has the additional knowledge required in order to identify the data subject?'

[102] See Joris van Hoboken, *Search engine freedom: on the implications of the right to freedom of expression for the legal governance of search engines* (Kluwer Law International 2012), p. 328.

*Situation (iv) – Hard for company and others to add a name*

In situation (iv), a behavioural targeting company processes data about an individual, and it would be difficult for *anybody* to tie a name to the data. Situation (iv) is probably rare in the behavioural targeting area, for it is often fairly easy for a behavioural targeting company to tie a name to the data it processes. Situation (iv) was discussed in Section 3. If the company uses the data to single out a person, the company processes personal data; it is not relevant whether a name can be tied to the data. As discussed in Section 6 below, the risk of harm remains, even if behavioural targeting companies do not tie a name to the data they hold about people.

## 6     The new Data Protection Regulation and pseudonymous data

The debate about the scope of the personal data definition was stirred up in 2012, when the European Commission published a proposal for a Data Protection Regulation that should replace the 1995 Directive.[103] The proposal did not bring any major changes to the 1995 personal data definition.[104] In the 2012 Commission proposal the absolute approach to identifiability was chosen. The personal data definition said that the 'means reasonably likely to be used by the controller *or by any other natural or legal person*' should be considered when determining identifiability.[105]

---

[103] COM(2012)11 of 25 Jan. 2012. I took part in this debate, for instance at the European Parliament (Frederik Zuiderveen Borgesius, Speech at the European Parliament: Interparliamentary Committee meeting: The reform of the EU Data Protection framework Building trust in a digital and global world (10 October 2012) <http://ssrn.com/abstract=2170174> accessed 24 January 2016.

[104] Art. 4(2) of the European Commission proposal for a Data Protection Regulation (COM(2012)11 of 25 January 2012). The proposed definition included 'online identifiers' and 'location data' in the list of examples of information by which somebody can be identified

[105] Ibid.,The proposed definition incorporates parts of Recital 26 of the Data Protection Directive ('the controller or by any other') in the definition of personal data. See also Recital 20 and 46 and Art. 3(2)(b) of the European Commission proposal for a Data Protection Regulation (COM(2012)11 of 25 January 2012). See also Impact Assessment for the proposal for a Data Protection Regulation (SEC(2012) 72 final of 25 Jan. 2012), p. 31.

The proposal led to much lobbying.[106] The Interactive Advertising Bureau, and companies such as Yahoo and Amazon (both using behavioural targeting), lobbied for a lighter regime for pseudonymous data: data about individuals without names attached.[107] Some European Parliament members proposed a lighter regime for pseudonymous data. [108] Other Parliament members proposed leaving pseudonymous data outside the scope of data protection law.[109]

In March 2014, the European Parliament adopted a compromise text, which the Parliament's LIBE Committee on Civil Liberties, Justice and Home Affairs prepared on the basis of the 3,999 amendments handed in by the members of parliament. [110] In the LIBE Compromise, personal data are defined roughly the same as in the 2012 proposal.[111] The preamble takes an absolute approach to identifiability.[112] Furthermore, in the LIBE Compromise a new category of personal data was introduced: 'pseudonymous data'.[113] In the LIBE Compromise, such pseudonymous data were subject to a lighter regime.[114]

In June 2015, the Council of the European Union published its proposal for the Regulation, to start negotiations with the European Parliament.[115] In the Council proposal, roughly the same personal data definition was used as in the Commission's 2012 proposal.[116] The Council proposal also included a definition of pseudonymisation, which resembles the pseudonymous data definition from the LIBE Compromise.[117]

In reaction to the Council proposal, the Working Party stressed that 'a natural person can be considered identifiable when, within a group of persons, [he or she] can be distinguished from others and consequently be treated differently. This means that the notion of identifiability should include singling out individuals.'[118] The Working Party was against the introduction of pseudonymous data as a new category of personal data.[119] Reding, when still a Euro Commissioner, warned that 'pseudonymous data must not become a Trojan horse at the heart of the Regulation, allowing the non-application of its provisions.'[120]

In December 2015, the Parliament and Council reached agreement on the data protection reform.[121] At the time of writing, the European Parliament and the Council must still formally adopt this final text. The final text of the Regulation defines personal data as follows:

> 'Personal data' means any information relating to an identified or identifiable natural person 'data subject'; an identifiable person is one who can be identified, directly or indirectly, in particular by reference to an identifier such as a name, an identification number, location data, online identifier or to one or more factors specific to the physical, physiological, genetic, mental, economic, cultural or social identity of that person. [122]

One of the main differences with the definition from the 1995 Data Protection Directive is that 'location data' and 'online identifier' are added as examples of identifiers.[123] The Regulation's preamble takes an absolute approach to identifiability, and uses the 'single out' phrase: 'To determine whether a person is identifiable, account should be taken of all the means reasonably likely to be used, *such as singling out, either by the controller or by any other person* to identify the individual directly or indirectly.'[124]

---

[120] Viviane Reding, The EU Data Protection Regulation: Promoting Technological Innovation and Safeguarding Citizens' Rights (Speech, Intervention at the Justice Council, 4 March 2014) <http://europa.eu/rapid/press-release_SPEECH-14-175_en.htm?locale=en> accessed 24 January 2016.
[121] Regulation (EU) No XXXX/2016 of the European Parliament and of the Council on the Protection of Individuals with Regard to the Processing of Personal Data and on the Free Movement of Such Data (General Data Protection Regulation), Consolidated text (outcome of the trilogue of 15/12/2015) <https://www.janalbrecht.eu/fileadmin/material/Dokumente/GDPR_consolidated_LIBE-vote-2015-12-17.pdf> accessed 24 January 2016.
[122] Art. 4(1) of the General Data Protection Regulation. Capitalisation adapted.
[123] The words 'genetic' and 'economic' (identity) are also new.

The final text of the Regulation also includes a definition of pseudonymisation:

> 'Pseudonymisation' means the processing of personal data in such a way that the data can no longer be attributed to a specific data subject without the use of additional information, as long as such additional information is kept separately and subject to technical and organisational measures to ensure non-attribution to an identified or identifiable person'.[125]

The Regulation's preamble says that pseudonymous data can still be personal data: 'Data which has undergone pseudonymisation, which could be attributed to a natural person by the use of additional information, should be considered as information on an identifiable natural person.'[126]

The final text of the Regulation treats pseudonymisation primarily as a data security measure.[127] For instance, the provision on data security gives pseudonymisation as an example of a security measure.[128] Apart from that, the Regulation suggests that, in some cases, companies that process pseudonymous data do not have to comply with data access requests from data subjects.[129] A discussion of that provision falls outside the scope of this paper.[130] The preamble of the LIBE Compromise suggested that companies did not have to obtain that data subject's prior consent for behavioural targeting based on pseudonymous data.[131] That suggestion is deleted in the final text of the Regulation.

---

[124] Recital 23 of the General Data Protection Regulation. Emphasis added.
[125] Art. 4(3b) of the General Data Protection Regulation. Capitalisation adapted.
[126] Recital 23 of the General Data Protection Regulation
[127] Art. 23(1); Art. 30(1)(a); Art. 83(1); Recital 60(a); Recital 61; Recital 67; Recital 125 of the General Data Protection Regulation.
[128] Art. 30(1)(a) of the General Data Protection Regulation.
[129] Art.10 of the General Data Protection Regulation. See also Recital 45.
[130] See on that topic: F.J. Zuiderveen Borgesius, *Improving Privacy Protection in the area of Behavioural Targeting* (Kluwer Law International 2015), p. 240-241.

In sum, the General Data Protection Regulation makes explicit that pseudonymous data can be personal data. While the preamble suggests that 'singling out' implies identifying somebody, the discussion about the exact scope of the personal data definition will probably continue.

## 7    Arguments in favour of the 'single out' interpretation

Many scholars say a logical interpretation of data protection law implies that data that are used to single out somebody should be seen as personal data.[132] But *why* should such data be regarded as personal data? We saw that it is often fairly easy to tie a name to nameless data.[133] Apart from that, there are four reasons why 'single out' data should be seen as personal data.

First, behavioural targeting triggers many concerns that lie at the core of data protection law. The risks of large-scale data collection are not overcome merely because data about a person are not tied to a name.[134] Harm can result from data processing, even if it is unlikely that a data subject's name can be tied to the data.

For instance, the massive collection of information about people's online activities may cause chilling effects: people adapt their behaviour if they know their activities may be tracked.[135] People who expect being monitored might hesitate to read about diseases, politics, or other topics. As Berners-Lee notes, 'we use the Internet to inform ourselves as voters in a democracy. We use the Internet to decide what is true and what is not. We use the Internet for healthcare and social interaction and so on.' He adds: 'These things will all have a completely different light cast on them if the users know that the click will be monitored and the data will be shared with third parties.'[136]

---

[131] Recital 38 and 58a of the LIBE Compromise. See on the consent requirement for personal data processing for behavioural targeting: Frederik Zuiderveen Borgesius, 'Personal data processing for behavioural targeting: which legal basis?' International Data Privacy Law (2015) (5-3) 163.
[132] See for instance Ronald Leenes, 'Do they know me? Deconstructing identifiability' University of Ottawa Law and Technology Journal (2008) 4(1-2) 135; Paul De Hert and Serge Gutwirth, 'Regulating profiling in a democratic constitutional state' in Mireille Hildebrandt and Serge Gutwirth (eds), *Profiling the European Citizen* (Springer 2008); Traung, *Op.cit*. But see for another view: Zwenne 2013, *Op.cit.*
[133] See section 5.
[134] Article 29 Working Party, 'Opinion 03/2013 on purpose limitation' (WP 203), 2 Apr. 2013, p. 46. See also Solon Barocas and Helen Nissenbaum, 'Big Data's End Run around Anonymity and Consent', in Julia Lane, Victoria Stodden, Stefan Bender, and Helen Nissenbaum (eds.), *Privacy, Big Data, and the Public Good: Frameworks for Engagement* (Cambridge University Press 2014).

In addition, with behavioural targeting, people lack control over information concerning them. People do not know which information about them is collected, how it is used, and with whom it is shared.[137] The feeling of lost control is a privacy problem.[138] Besides, storing personal information about millions of people is inherently risky. A data breach could occur, or data could be used for unexpected purposes. For example, commercial databases tend to attract the attention of law enforcement bodies.[139] The police often request data from companies like Facebook and Google, both using behavioural targeting.[140] Moreover, intelligence agencies could access data held by companies, or identify people by exploiting commercial tracking cookies.[141] Schneier summarises: 'the primary business model of the Internet is built on mass surveillance, and our government's intelligence-gathering agencies have become addicted to that data.'[142]

---

[135] See on chilling effects in the context of behavioural targeting: Aleecia M McDonald AM and Lorrie F. Cranor, 'Beliefs and Behaviors: Internet Users' Understanding of Behavioral Advertising (38th Research Conference on Communication, Information and Internet Policy, Telecommunications Policy Research Conference) (2 October 2010) <http://ssrn.com/abstract=1989092> accessed 24 January 2016. Another study analysed Google search results, and suggests people 'were less likely to search using search terms that they believed might get them in trouble with the U. S. government' (Alex Marthews and Catherine Tucker, 'Government Surveillance and Internet Search Behavior' (March 2014) <http://ssrn.com/abstract=2412564> accessed 24 January 2016).

[136] Tim Berners-Lee, 'No Snooping' (3 November 2009) <www.w3.org/DesignIssues/NoSnooping.html> accessed 24 January 2016 (apparent typo corrected; 'them' instead of 'then').

[137] Alessandro Acquisti and Jens Grossklags, 'What can behavioral economics teach us about privacy?' in Alessandro Acquisti, Stefanos Gritzalis, Costos Lambrinoudakis, Sabrina di Vimercati (eds), *Digital Privacy: Theory, Technologies and Practices* (Auerbach Publications, Taylor and Francis Group 2007).

[138] See: Ryan Calo, 'The boundaries of privacy harm' Indiana Law Journal (2011)86 1131.

[139] See on state access to commercial data: Ian Brown, 'Government access to private-sector data in the United Kingdom' (2012) 2(4) International Data Privacy Law 230; Axel Arnbak, Joris Van Hoboken, and Nico Van Eijk, 'Obscured by Clouds or How to Address Governmental Access to Cloud Data from Abroad' (2013) <http://ssrn.com/abstract=2276103> accessed 24 January 2016.

[140] See Google Transparency Report <www.google.com/transparencyreport/>; Facebook Global Government Requests Report <www.facebook.com/about/government_requests> accessed 24 January 2016.

Furthermore, marketing companies compile detailed information about people and can classify people. Behavioural targeting data could be used for discriminatory practices. As Turow notes, companies can sort people into 'targets' and 'waste', and treat them as such.[143] For instance, some companies categorise people (cookies) as 'disabled/handicapped',[144] or as 'lesbian, gay, bisexual, transgender'.[145] In sum, the online marketing industry processes large amounts of information about millions of people, and this involves risks, also when companies do not tie a name to data they store about individuals.

A second reason why data used to single out a person should be seen as personal data is that a name is merely one of the identifiers that can be tied to data about a person – and not even the most effective identifier. In some situations, a name is the most practical identifier. If a company wanted to tie a profile based on information gathered through a loyalty card of a supermarket to an online profile, it would help if a name were linked to both profiles. But for many purposes, a name is not the most effective identifier. If a company wants to send messages to a phone, or track its location, a phone number or another phone ID is the easiest identifier. For an ad network that wants to track somebody's online behaviour or target ads to a person, a cookie is a better identifier than a name. Many companies are not interested in tying a name to data they process for behavioural targeting, even though they could easily do so. Furthermore, a unique number is often a better identifier than a name, because names may not be unique.[146]

[141] See Morgan Marquis-Boire, Glenn Greenwald and Micah Lee, 'XKEYSCORE: NSA's Google For The World's Private Communications' (The Intercept, 1 July 2015) <https://firstlook.org/theintercept/2015/07/01/nsas-google-worlds-private-communications/> accessed 24 January 2016.
[142] Bruce Schneier, 'The public-private surveillance partnership' (Bloomberg, 31 July 2013) <www.bloombergview.com/articles/2013-07-31/the-public-private-surveillance-partnership> accessed 24 January 2016. He is from the US, but his remarks are relevant for Europe too.
[143] Joseph Turow, *The Daily You: How the New Advertising Industry is Defining Your Identity and Your Worth* (Yale University Press 2011), chapter 4.
[144] Rocket Fuel, Health Related Segments 2014 <http://rocketfuel.com/downloads/Rocket%20Fuel%20Health%20Segments.pdf> accessed 24 January 2016.
[145] Flurry Audiences, Segment audiences by real interests <www.flurry.com/flurry-personas.html> accessed 5 October 2013. On 24 January 2016, the transgender category was not listed anymore on the site.

Third, the whole point of behavioural targeting is tracking individuals, building profiles of individuals, and targeting ads to individuals. The goal of behavioural targeting is, in the words of a marketing company, 'to use data to deliver the right ad to the right person at the right time.'[147] To do this, behavioural targeting companies must single out people with unique identifiers. As Turow notes, 'if a company can follow your behavior in the digital environment – an environment that potentially includes your mobile phone and television set – its claim that you are 'anonymous' is meaningless. (…) It matters little whether your name is John Smith, Yesh Mispar, or 3211466.'[148] The International Working Group on Data Protection in Telecommunications makes a similar point: 'While ads may well be addressed to a machine at the technical level, it is not the machine which in the end buys the proverbial beautiful pair of red shoes – it is an individual.'[149] In sum, targeted advertising aims to influence individuals. Therefore, it makes sense to apply the data protection regime, which was developed to protect fairness and fundamental rights, such as privacy, when information about individuals is processed.[150]

Fourth, seeing data that are used to single out a person as personal data fits the rationale for data protection law. The Court of Justice of the European

---

[146] The Working Party notes that very common names by itself are not always personal data, because they cannot be used to identify people (Article 29 Working Party 2007, WP 136, *Op.cit.*, p.13).

[147] Brian Lesser, 'How to Use Data to Deliver the Right Ad to the Right Person at the Right Time', AdAge 3 July 2012, <http://adage.com/article/digitalnext/data-deliver-ad-person-time/235734/> accessed 24 January 2016.

[148] Turow 2011, *Op. Cit.*, p. 7.

[149] International Working Group on Data Protection in Telecommunications 2013, *Op.cit.* p. 3. This Berlin Group was founded in 1983 and consists of representatives from Data Protection Authorities and other bodies of national public administrations, international organisations and scientists from around the world.

[150] Art. 1(1) and Art. 6(1)(a) of The Data Protection Directive. A second goal of the Data Protection Directive is ensuring the free flow of data within the EU (Art. 1(2) of The Data Protection Directive).

Union says the Directive aims for a 'high level' of protection,[151] and that fundamental rights guide the interpretation of the Directive.[152] Furthermore, 'limitations in relation to the protection of personal data must apply only in so far as is strictly necessary.'[153]

According to the European Court of Human Rights, the right to private life is a broad term that should be applied in a 'dynamic and evolutive' manner.[154] The right to private life 'must be interpreted in such a way as to guarantee not rights that are theoretical or illusory but rights that are practical and effective.'[155] The Court says that 'information derived from the monitoring of personal Internet usage' is also protected by the right to private life.[156]

I argue that, like the European Convention on Human Rights, data protection law should be applied pragmatically and dynamically.[157] In the light of new developments, such as behavioural targeting, 'big data', and the 'internet of things', the scope of data protection law should not be limited to data that can be tied to people's names.

A fifth reason to regard nameless individual behavioural targeting profiles as personal data is the fact that it is often fairly easy to tie a name to the data, as discussed in Section 5. In sum, data protection law should apply to information that is used to single out people, even if no name can be tied to the information.

---

[151] Case C-524/06, *Huber*, EU:C:2008:724, par. 50; Case C-131/12, *Google Spain*, EU:C:2014:317, par. 66.
[152] Case C-465/00, C-138/01 and C-139/01, *Österreichischer Rundfunk*, EU:C:2003:294, par. 68; Case C-131/12, Google Spain, EU:C:2014:317, par. 68.
[153] See e.g. Case C-293/12 and C-594/12, *Digital Rights Ireland Ltd, EU:C:2014:238*, par. 52; Case C-473/12, *Institut professionnel des agents immobiliers,* EU:C:2013:715, par. 39 (with further references).
[154] ECtHR, Christine Goodwin v. United Kingdom, Appl. No. 28957/95, 11 July 2002, par 74.
[155] ECtHR, Armonas v. Lithuania, Appl. No. 36919/02, 25 November 2008, par. 38.
[156] ECtHR, Copland v. United Kingdom, No. 62617/00, 3 April 2007, par. 41
[157] The ECtHR connects data protection law and Art. 8 of the ECHR: '[t]hat broad interpretation [of the right to private life of Art 8 ECHR] corresponds with that of the Council of Europe's Convention of 28 Jan. 1981 for the Protection of Individuals with regard to Automatic Processing of Personal Data (…)' (ECtHR, Amann v. Switzerland, Appl. No. 27798/95, 16 February 2000, par. 65). See similarly: ECtHR, Rotaru v. Romania, Appl. No. 28341/95, 4 May 2000, par. 43.

## 8    Arguments against the 'single out' interpretation

Commentators have brought forward several arguments against the 'single out' interpretation of personal data.[158] Four of the main arguments are summarised here. I conclude that they are not persuasive.

First, it has been argued that companies will have fewer incentives to pseudonymise data, if data protection law applies to 'single out data' or pseudonymised data.[159] This may be true, but when information is within the scope of data protection law, companies still have an incentive to pseudonymise data. Data protection law requires appropriate security when personal data are processed, and pseudonymisation can improve security.[160] For instance, pseudonymisation can help to keep data subjects' names hidden for employees who do not need to see the names. Pseudonymisation also mitigates the risks resulting from a data breach. Take the case of a behavioural targeting company that suffers a data breach: nameless individual profiles regarding people's browsing behaviour are accidentally published on the web. People who see the profiles learn that the person behind the cookie with ID *xyz* visited www.embarrassing-website.com, but they do not see the name of this person immediately. Hence, the privacy risks are mitigated, because the data breach concerns nameless data. There is less risk of embarrassment or other unpleasant surprises for the person behind the cookie with ID *xyz*. We saw, however, that it is often possible to find the name of the person behind a nameless profile.[161] The security requirements of data protection law are an incentive to pseudonymise data, but replacing a name with another unique identifier

[158] Zwenne makes the most detailed and eloquent argument against a broad interpretation of personal data (Zwenne 2013, *Op.cit.*). See also Stevens 2015, *Op.cit*, p. 105.
[159] G.J. Zwenne, Over Persoonsgegevens en IP-adressen, en de Toekomst van Privacywetgeving [On Personal Data and IP addresses, and the Future of Privacy Legislation] in Mommers et al. (eds), *Het Binnenste Buiten. Liber Amicorum ter Gelegenheid van het Emiritaat van Prof. dr. Aernout H.J. Schmidt, Hoogleraar Recht en Informatica te Leiden [The Inside Out. Liber Amicorum for Retirement of Prof. Dr. Aernout H. J. Schmidt, Professor of Law and Computer Science in Leiden]* (eLaw@Leiden 2010), p. 336.
[160] See Art. 17 of the Data Protection Directive; Article 29 Working Party 2014, WP 216, *Op.cit.*
[161] See section 5.

does not render data anonymous or keep data outside the scope of data protection law.[162]

Second, some suggest that applying data protection law to behavioural targeting data would be bad for the economy and innovation.[163] Even if this were true, the argument would not be sufficient to keep nameless behavioural targeting data outside the scope of data protection law. Processing is not prohibited when information is within the scope of data protection law. Nevertheless, some companies might make less profit when they must comply with data protection law. Obtaining legal advice or complying with legal data security requirements is costly, for instance. However, even if fundamental rights were ignored and only economic effects were considered, the most relevant question would be whether society as a whole gains or loses.

From an economic perspective, it is unclear whether more or less legal protection of personal data is better. As Acquisti, the leading scholar on the economics of privacy, puts it, 'economic theory shows that, depending on conditions and assumptions, the protection of personal privacy can increase aggregate welfare as much as the interruption of data flows can decrease it.'[164]

Furthermore, although innovation (a term almost as vague as privacy) and economic growth are important, they do not trump fundamental rights. We should not allow children younger than eight being employed in factories, even if innovation or the economy would benefit from it.[165] And it would also lead to innovation if regulation pushes companies towards developing new privacy preserving behavioural targeting technologies.[166]

---

[162] See European Agency for Fundamental Rights 2014, *Handbook on European data protection law, first edition* (Publications Office of the European Union 2014), pp. 45-46; Article 29 Working Party 2014, WP 216, *Op.cit.*, p. 20.
[163] See e.g. Stringer, *Op.cit.*
[164] Alessandro Acquisti, 'The economics of personal data and the economics of privacy' (background paper conference: The Economics of Personal Data and Privacy: 30 Years after the OECD Privacy Guidelines) (2010) <www.oecd.org/internet/ieconomy/46968784.pdf> accessed 24 January 2016, p. 19. See along similar lines: Alessandro Acquisti, 'The Economics and Behavioral Economics of Privacy' in Julia Lane, Victoria Stodden, Stefan Bender, and Helen Nissenbaum (eds.), *Privacy, Big Data, and the Public Good: Frameworks for Engagement* (Cambridge University Press 2014), p. 90.
[165] Helen Nissenbaum made a remark among these lines at the Acatech Symposium Internet Privacy (26 March 2012, Berlin). Art. 32 of the Charter of Fundamental rights of the European Union says: 'The employment of children is prohibited.'

Third, a broad interpretation of personal data implies that data protection law applies even when there are no privacy threats. Some suggest that data protection law should not be severed from the right to privacy.[167] This argument does not fit in well with positive law, as the Charter of Fundamental Rights of the European Union distinguishes between the right to data protection and the right to privacy.[168] Moreover, many authors say it is an advantage that data protection law applies to personal data rather than only to private data.[169] After all, data protection law is not merely aimed at protecting privacy but also at achieving fairness more generally.[170]

Fourth, some worry that almost everything could become personal data if 'single out' data were seen as personal data. Enforcement of data protection law would become too difficult. Data Protection Authorities would only be able to enforce the law against some wrongdoers. This could lead to arbitrary decisions about enforcement, which would be detrimental to legal certainty.[171] Additionally, the scope of the personal data definition would become too fuzzy, which could also harm legal certainty.

The argument that the broad scope of data protection law makes enforcement difficult has some merit. But excluding 'single out' data from the scope of data protection law altogether would not be the right reaction.

---

[166] See Acquisti 2010, *Op.cit.*

[167] See Colette Cuijpers and Paul Marcelis, 'Oprekking van het Concept Persoonsgegevens Beperking van Privacybescherming?' [Stretching the Concept of Personal Data, Limiting the Protection of Privacy?] (2012)(6) Computerrecht 339-351.

[168] See Art. 7 and 8 of the Charter of Fundamental Rights of the European Union. See on the difference between the right to privacy and the right to protection of personal data: Gloria González Fuster, *The Emergence of Personal Data Protection as a Fundamental Right of the EU* (Springer 2014); Raphael Gellert and Serge Gutwirth, 'The Legal Construction of Privacy and Data Protection' [2013] 29 Computer Law & Security Review 522; Gloria González Fuster and Serge Gutwirth, 'Opening up Personal Data Protection: A Conceptual Controversy' [2013] 29 Computer Law & Security Review 531.

[169] See e.g. Paul De Hert and Serge Gutwirth, Privacy, Data Protection and Law Enforcement. Opacity of the Individual and Transparency of Power in Eric Claes, Anthony Duff and Serge Gutwirth (eds), *Privacy and the Criminal Law* (Intersentia 2006), p. 94.

[170] See Art. 6(1)(a) of the Data Protection Directive. See also Art. 5(1)(a) of the Data Protection Regulation (final text).

[171] See Zwenne 2013, *Op.cit.*, in particular pp. 33-35.

By the same token, it is appropriate that we have environmental law, even though it is impossible to catch every polluter. Besides, in legal practice the fringes of a definition can always provoke discussion.

In sum, the counter-arguments do not justify leaving behavioural targeting outside the scope of data protection law. Of course, merely ensuring that data protection law applies to behavioural targeting does not solve the privacy problems of behavioural targeting. Data protection law has weaknesses, and compliance and enforcement are often lacking.[172] But at least data protection law provides a framework to assess fairness when information about individuals is used.[173] And data protection law can help to make personal data processing transparent, as it requires companies to disclose information about their processing practices.[174] If problems are found, this could warrant the conclusion that regulatory intervention is needed.

## 9    Conclusion

Two conclusions can be drawn. First, data protection law generally applies to behavioural targeting. European Data Protection Authorities say that a company processes personal data if it uses data to single out a person, regardless of whether a name can be tied to the data. Second, from a normative perspective, data protection law should apply. Behavioural targeting triggers many concerns that lie at the core of data protection law. The risks of large-scale data collection are not overcome merely because data about a person are not tied to a name. The whole point of behavioural targeting is singling out *individuals*: tracking individuals, profiling individuals, and targeting individuals. Furthermore, a name is merely one of the identifiers that can be tied to data about a person, and it is not even the most practical identifier for behavioural targeting. Seeing data used to

---

[172] See Zuiderveen Borgesius 2015, chapter 8 section 2, *Op.cit.*
[173] See in particular Art. 6 and 7 of the Data Protection Directive.
[174] See Art. 10 and 11 of the Data Protection Directive. See also CJEU, Case C-201/14, Smaranda Bara and Others v Președintele Casei Naționale de Asigurări de Sănătate and Others, 1 October 2015.

*single out* a person as personal data fits the rationale for data protection law: protecting fairness and fundamental rights. Lastly, it is often fairly easy for companies to tie a name to nameless individual profiles. In sum, data that are used to single out a person should be considered personal data.

* * *